\RequirePackage[final]{graphicx}
\documentclass[aps, superscriptaddress,pra,twocolumn,footinbib]{revtex4-1}
\usepackage[utf8]{inputenc}
\usepackage{amsmath}
\usepackage{amsfonts}
\usepackage{mathtools}
\usepackage{graphicx}
\usepackage{fixme}
\usepackage{color}

\def\bk{{\bf k}}
\def\bp{{\bf p}}

\begin{document}

\title{Time-reversal invariant topological superfluids in Bose-Fermi mixtures}
\date{May 2017}

\author{Jonatan Melk\ae r \surname{Midtgaard}}
\affiliation{Department of Physics and Astronomy,  Aarhus University, Ny Munkegade, DK-8000 Aarhus C, Denmark}
\author{Zhigang Wu}
\affiliation{Institute for Advanced Study, Tsinghua University, Beijing, 100084, China}
\author{G.\ M.\ Bruun}
\affiliation{Department of Physics and Astronomy,  Aarhus University, Ny Munkegade, DK-8000 Aarhus C, Denmark}

\begin{abstract}
A mixed dimensional system of fermions in two layers immersed in a  Bose-Einstein condensate (BEC) is shown to be a promising setup  to realise topological superfluids with 
time-reversal symmetry (TRS). The induced  interaction between the fermions mediated by the BEC gives rise to a
 competition between $p$-wave pairing within each layer and $s$-wave pairing between the layers. When the layers are far apart,  intra-layer pairing dominates and the system forms a topological superfluid either with or without TRS. With decreasing layer separation or increasing BEC coherence length, inter-layer pairing sets in. We show that this  
leads either to a second order transition breaking TRS where the edge modes gradually become gapped, or to a first order transition to a topologically trivial $s$-wave 
superfluid. Our results provide a realistic roadmap for experimentally realising a topological superfluid with TRS in a cold atomic system. 
\end{abstract}

\maketitle

\section{Introduction}
The  search for  superfluids/superconductors with non-trivial topological properties  has experienced an explosion of activities in recent years. 
One reason is that these systems can host gapless edge (Majorana) modes with possible applications in  quantum computation~\cite{Alicea2012,Nayak2008}.
Excitingly, evidence for topological superconductivity and gapless edge states have been 
reported in
nano-wires~\cite{Mourik2012,Deng2012,Das2012,Rokhinson2012,Finck2013,NadjPerge2014,Albrecht2016}.
So far the focus has predominantly been placed on superfluids~\cite{footnote} without TRS, which belong to the symmetry 
class D in the 10-fold classification scheme of topological insulators/superfluids~\cite{Schnyder2008,Kitaev2009,Altland1997}.  However, 
 superfluids with TRS, belonging to the class DIII, can 
also host gapless Majorana mode pairs, which are protected by Kramers theorem. There are also several 
proposals to realise such systems in laboratory, both in condensed 
matter systems~\cite{Wong2012,Deng2012b,Zhang2013,Nakosai2013,Keselman2013,Gaidamauskas2014,Klinovaja2014,Schrade2015} and in 
 cold atomic systems~\cite{Yan2013,Yang2013,Huang2015}.  
One example of such intriguing systems is the superfluid $^3$He B phase, whose topological properties have been studied recently~\cite{Volovik2009,Volovik2014}. 
 However, one has yet to observe a topological superfluid with TRS in a cold atomic system.

Recently, we showed that a mixed dimensional atomic gas system consisting of a two-dimensional (2D) layer of  fermions immersed in a 3D BEC 
constitutes a  promising system for realising a $\mathbb{Z}$ topological superfluid in class D with a high critical temperature~\cite{Wu2016,Midtgaard2016}.  Here, we show that an analogous system with two layers of fermions, first studied in Ref.~\cite{Nishida2009}, is 
naturally suited to 
 realise a  $\mathbb{Z}_2$ topological superfluid with TRS. Fermions in the  layers interact attractively via an induced interaction 
 mediated by the BEC. The relative strengths of the intra- and inter-layer induced interaction results in a  competition between  $p_x\pm ip_y$-wave pairing involving
 fermions in the same layer, and  $s$-wave pairing involving fermions in different layers. For large distance between the layers,  intra-layer pairing
 dominates and one has either a $(p_x+ ip_y)\times(p_x-ip_y)$  system with TRS or a $(p_x+ ip_y)\times(p_x+ip_y)$ without TRS. With decreasing layer distance or increasing 
  BEC coherence length, we show that
   inter-layer $s$-wave pairing occurs in a second order transition for the $(p_x+ ip_y)\times(p_x-ip_y)$ system, 
 which breaks TRS thereby gradually gapping the  edge modes  
 without closing the bulk gap. For short layer distance, the system ends up in a topologically trivial $s$-wave superfluid, resembling the case of a single layer with two spin components~\cite{Anzai2017}. On the other hand, the transition from the topological 
 $(p_x+ ip_y)\times(p_x+ip_y)$ to the trivial $s$-wave superfluid is of the first order.

\begin{figure}[htb]
\centering
\includegraphics[width=\columnwidth]{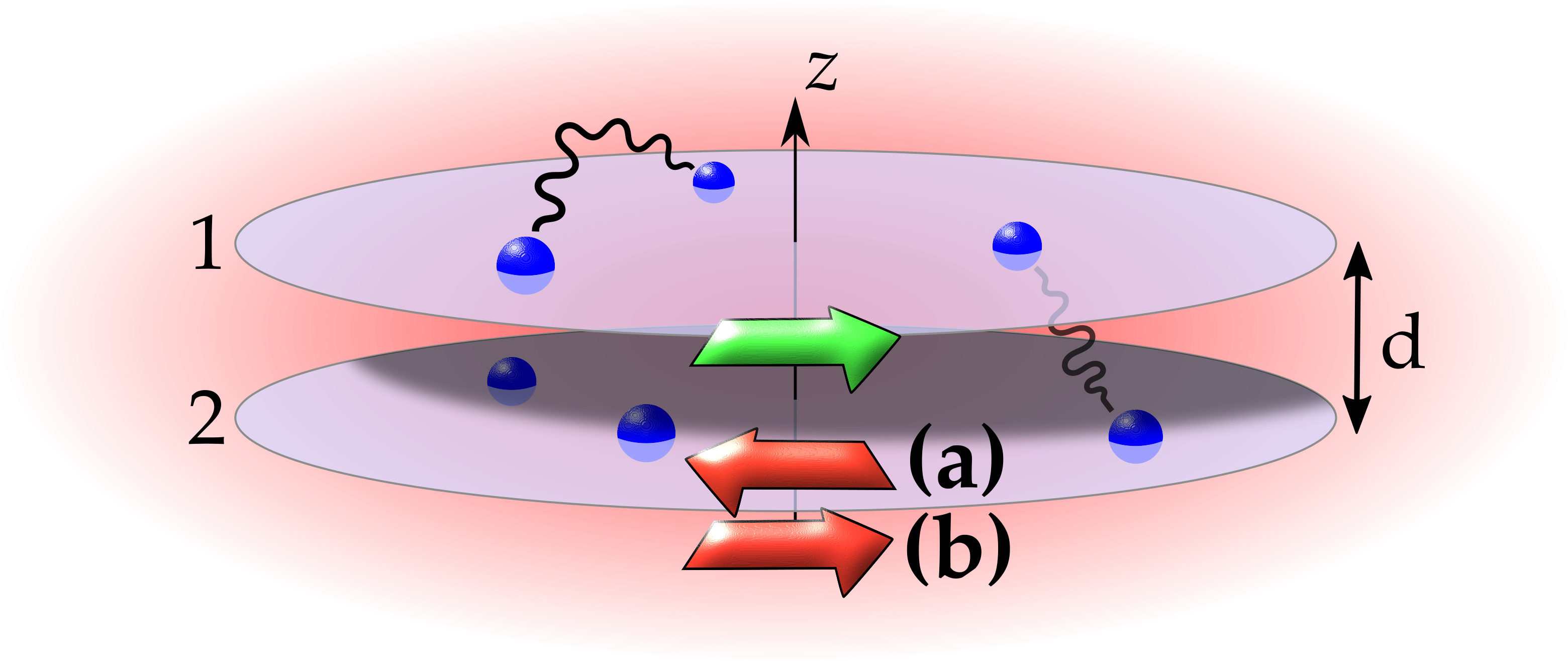}
\caption{(Color online). In the proposed experimental setup, fermions (blue spheres) confined to two  layers with distance $d$, interact with the surrounding BEC (red background). This results 
in induced intra-layer and inter-layer interactions (illustrated by black wiggly lines). The green and red arrows indicate the  
edge modes in the two layers respectively. The intra-layer p-wave pairings are either of \textbf{(a)} different chirality, realizing a $\mathbb{Z}_2$ topological superfluid, or of \textbf{(b)} same chirality, realizing a $\mathbb{Z}$ topological superfluid.
}
\label{fig:system}
\end{figure}

\section{Model}
We consider identical (spin polarised) fermions of mass $m$ 
 in two layers located at $z=0$ and $z=d$, see Fig.\ \ref{fig:system}. 
The fermions are immersed in a 3D gas of bosons with mass $m_B$ and density $n_B$. The partition function of the system at temperature $T$ is 
\begin{align}
\mathcal{Z} = \int \mathcal{D}(\bar{\psi}_F,\psi_F,\psi_B^\ast,\psi_B) \, e^{-\left(S_\text{F}+S_\text{B}+S_\text{int}\right)},
\label{eq:action}
\end{align}
where $\psi_B({\mathbf r},\tau)$ and $\psi_F({\mathbf r},\tau)$ are the bosonic and fermionic fields at point ${\mathbf r}$ and imaginary time $\tau$.
The bosons form a  weakly interacting BEC that be described by  Bogoliubov theory, which  yields
\begin{align}
S_B = \beta\sum_{{\mathbf p}\neq 0,l}\gamma_p^{\ast}(-i\omega_l+E_{\mathbf p})\gamma_p
\end{align}
for the bosonic part of the action, where $\beta = 1/T$, 
 $\omega_l =2l\pi T$ with $l=0,\pm 1,\pm 2, \ldots$ 
are the Bose Matsubara frequencies and $\gamma_p$ describes the quasi-particle with momentum $\bp=(p_x,p_y,p_z)$ and energy $E_\bp$. Here we have defined $p\equiv (\mathbf p,i\omega_l)$. The Bogoliubov spectrum is 
$E_{\mathbf p} =\sqrt{ \epsilon_{\mathbf p} (\epsilon_{\mathbf q} + 2 g_Bn_B) }$, where  $\epsilon_{\mathbf p} = \mathbf p^2/2m_B$ and $g_B = 4\pi a_B/m_B$, where $a_B$ is the boson scattering length. The fermion part of the action is
\begin{align}
S_F = \beta\sum_{\sigma}\sum_{{\mathbf k}_\perp,j}\bar a_{k_\perp\sigma}(-i\omega_j+\xi_{\mathbf k_\perp})a_{k_\perp\sigma}
\end{align}
where $a_{k_\perp\sigma}$ are the Grassmann fields for the fermions in layer $\sigma=1,2$. The effective 2D action for the fermions 
results from the fact that the vertical trapping potentials are sufficiently tight that the fermions reside only in the 
lowest trap levels $\phi_0(z)$ and $\phi_0(z-d)$ along the $z$-direction. We have defined 
$k_\perp\equiv (\mathbf k_\perp,i\omega_j)$ with $\mathbf k_\perp=(k_x,k_y)$ as the in-plane momentum,
 $\omega_j =(2j+1)\pi T$ with $j=0,\pm 1,\pm 2, \ldots$ 
are the Fermi Matsubara frequencies, and $\xi_{\mathbf k_\perp}=\mathbf k_\perp^2/2m-\mu$ where $\mu$ is the chemical potential of the fermions.
We take $\mu$ to be the same in each layer which contains an equal number of fermions. Finally, the Bose-Fermi interaction is 
\begin{align}
S_\text{int} = g\int\! d^3r\int_0^\beta\! d\tau\,\bar{\psi}_F\psi_F\psi_B^\ast\psi_B,
\end{align}
 where $g$ is the boson-fermion interaction strength. Using the Bogoliubov theory to write 
$\psi_B({\mathbf r},\tau)=\mathcal V^{-1/2}\sum_p(u_\mathbf p \gamma_p - v_\mathbf p \gamma^*_{-p})\exp[i(\mathbf p\cdot\mathbf r-\omega_l\tau)]$ with 
$u_{\mathbf p}^2, v_{\mathbf p}^2 =[ ({\epsilon_{\mathbf p} + g_Bn_B})/{E_{\mathbf p}} \pm 1 ]/2$, and expanding the fermonic fields as $\psi_F({\mathbf r},\tau)=\sum_{p_\perp,\sigma}a_{p_\perp\sigma}\exp[i(\mathbf p\cdot\mathbf r_\perp-\omega_j\tau)\phi_0(z-(\sigma-1) d)/\sqrt\mathcal A$, we find 
\begin{gather}
S_\text{int} = 
\frac gT\sqrt{\frac{n_B}{\mathcal V}}\sum_{\substack{{\mathbf p}\neq 0\\l,\sigma}}\sqrt{\frac{\epsilon_{\mathbf p}}{E_{\mathbf p}}}
(\gamma_p+\gamma_p^\ast)\rho_{p_\perp\sigma}e^{-ip_zd(\sigma-1)}
\label{Interaction}
\end{gather}
where $\mathcal V$ is the BEC volume, $\mathcal A$ is the area of the Fermi layer, $\rho_{p_\perp\sigma}=\sum_{k_\perp}\bar a_{k_\perp-p_\perp\sigma}a_{k_\perp\sigma}$ 
 and  $p_\perp\equiv(\mathbf{p_\perp},i\omega_l)$. 
 
Integrating out the quadratic Bose fields in the action in (\ref{eq:action}) yields the effective action 
\begin{align}
S_\text{eff}=S_F+ \frac{\beta}{2\mathcal{A}}  \sum_{\substack{p_\perp\\\sigma,\sigma'}} {\rho}_{-p_\perp\sigma}V_\text{ind}^{\sigma\sigma'}( p_\perp)
  \rho_{p_\perp\sigma'},
\end{align}
where the induced interaction between the fermions, mediated by the bosons, is 
\begin{align}
V^{\sigma\sigma'}_\text{ind}(p_\perp) = g^2 \int\! \frac{dp_z}{2\pi} e^{ip_zd(\sigma-\sigma')} \chi_\text{BEC}(p).
\label{Vinduced}
\end{align}
Here, $ \chi_\text{BEC}(p) = n_B\mathbf p^2m_B^{-1}/[(i\omega_l)^2-E_\mathbf{p}^2]$
is the density-density correlation function for the BEC and the $p_z$-integration in (\ref{Vinduced}) is due to the fact that the momentum along the $z$-direction is not conserved  in the
boson-fermion scattering due to the mixed dimensional setup. We note that the induced interaction in Eq.~(\ref{Vinduced}) is obtained with the assumption that the 3D BEC is not affected by the 2D Fermi gases. This is justified in our mixed dimensional setup because the properties of the 3D BEC will only be affected locally in the vicinity of the 2D layers. Since the induced interaction between the fermions is determined by the overall bulk properties of the BEC, we expect that these local effects on the 3D BEC will only lead to small corrections to the induced interaction given by Eq.~(\ref{Vinduced}). For zero frequency, $i\omega_l= 0$, performing the $p_z$ integrals yields 
\begin{equation}
V^{\sigma\sigma'}_\text{ind}(\mathbf{p}_\perp,0)  = - \frac{2g^2 n_B m_B}{\sqrt{\mathbf p_\perp^2 + 2/\xi_B^2}}
e^{-d|\sigma-\sigma'| \sqrt{\mathbf p_\perp^2 + 2/\xi_B^2} },
\label{InducedInt}
\end{equation}
where $\xi_B=(8\pi n_B a_B)^{-1/2}$ is the BEC coherence length. 
The inter-layer ($\sigma\neq\sigma'$) interaction is suppressed 
compared to the intra-layer ($\sigma=\sigma'$) interaction by an exponential factor related to the layer distance $d$.
Fourier transforming (\ref{InducedInt}) yields a Yukawa interaction $V(\mathbf r) = -g^2n_B m_B\pi^{-1} \exp(-\sqrt{2}r/\xi_B)/r$ in real space 
with a range determined by $\xi_B$~\cite{Bijlsma2000,Viverit2000,Midtgaard2016,Suchet2017}. Here $r=|\mathbf r|$ is the distance between the 
particles, which can reside in the same or in different planes.

\section{Gap equations}
Since the induced interaction given by (\ref{InducedInt}) is attractive,  fermions with opposite momenta can form Cooper pairs 
within each layer (intra-layer pairing) as well as between  different layers (inter-layer pairing). The BCS Hamiltonian 
describing such parings is   
\begin{equation}
H_\text{BCS} = \frac{1}{2} \sum_{\mathbf{p}}
\Psi^\dagger(\mathbf{p})
\mathcal{H}(\mathbf{p})
\Psi(\mathbf{p}),
\label{HBCS}
\end{equation}
where $\Psi(\mathbf{p}) = (a_{\mathbf{p} 1}, a^{\dagger}_{\mathbf{-p} 1}, a_{\mathbf{p} 2}, a^{\dagger}_{\mathbf{-p} 2})^T$ and
\begin{equation}
\mathcal{H}(\mathbf{p}) =
\begin{bmatrix}
\xi_\mathbf{p} & \Delta_{11}(\mathbf{p}) & 0 &\Delta_{12}(\mathbf{p})   \\
\Delta^\ast_{11}(\mathbf{p})& -\xi_\mathbf{p} & -\Delta^\ast_{12}(\mathbf{p}) & 0 \\
0 & -\Delta_{12}(\mathbf{p}) & \xi_\mathbf{p} & \Delta_{22}(\mathbf{p}) \\
\Delta^\ast_{12}(\mathbf{p}) & 0 & \Delta^\ast_{22}(\mathbf{p}) & -\xi_\mathbf{p}
\end{bmatrix}.
\label{eq:BdGham}
\end{equation}
Here the $\perp$-subscript is dropped since we are dealing only with  2D momenta of the fermions from now on, and
 $a_{\mathbf{p}\sigma}$ are the  fermi annihilation operators for  layer $\sigma=1,2$. 
We neglect retardation effects and use only the zero frequency component of the induced interaction. Retardation effects    
are small when the Fermi velocity $v_F$ in the layers is much smaller than the speed of sound in the BEC, while for larger $v_F$ they 
 suppress the magnitude of the pairing without changing the qualitative behavior~\cite{Wu2016}. 
The pairing fields are determined self-consistently as  
\begin{equation}
\Delta_{\sigma\sigma'}(\mathbf{p}) = -\sum_{\mathbf{k}} V^{\sigma\sigma'}_\text{ind}(\mathbf{p}-\mathbf{k},0)\langle a_{\mathbf{k}\sigma}  a_{\mathbf{-k}\sigma'} \rangle.
\label{Deltadef}
\end{equation}
We take the inter-layer pairing to be $s$-wave so that  $\Delta_{12}(\mathbf{p}) = \Delta_{12}(-\mathbf{p})=  -\Delta_{21}(\mathbf{p})$ and the  
Fermi anti-symmetry dictates that  $\Delta_{\sigma\sigma}(\mathbf{p})=-\Delta_{\sigma\sigma}(-\mathbf{p})$ for the intra-layer 
pairing. Since the system has rotational symmetry with respect to the $z$-axis, we take the intra-layer pairing to be of the     
 $p_x\pm ip_y$ form, as this fully gaps the Fermi surface~\cite{Anderson1961}, i.e.\ $\Delta_{\sigma\sigma}(\mathbf{p})=\Delta_\sigma(|\mathbf{p}|)e^{i\phi_\sigma(\bp)}$ 
 where  $\phi_\sigma(\bp) = \phi_{0\sigma} \pm \varphi_\bp $ with $\varphi_\bp $ being the  azimuthal angle  of $\mathbf p$. Furthermore, for identical layers we assume that $\Delta_{1}(|\bp|) = \Delta_{2}(|\bp|)$ and we thus have
$ \Delta_{22}(\bp)= \Delta_{11}(\bp)  e^{i\left[\phi_2(\bp) - \phi_1(\bp) \right ]}$. 
We diagonalise (\ref{HBCS})
by introducing new pairing fields $\Delta_{\pm}(\bp) = \Delta_{11}(\bp) \pm  \Delta_{12}(\bp) e^{-i\left[\phi_2(\bp)- \phi_1(\bp)-\pi\right ]/2}$. 
Equation (\ref{Deltadef}) then yields 
a set of gap equations in a symmetrical form as
\begin{align}
\Delta_{\nu}(\mathbf{p}) &= -\sum_{\nu',\mathbf{k}} V_{\nu\nu'}(\mathbf{p}-\mathbf{k}) \frac{\Delta_{\nu'}(\bk)}{2E_{\bk,\nu'}} \tanh\left(\frac{E_{\bk,\nu'}}{2T}\right). \label{gap3}
\end{align}
Here $\nu=\pm$, $E_{\bp,\pm} = \sqrt{\xi_\bp^2 + |\Delta_{\pm}(\bp) |^2}$, and 
\begin{align}
 V_{\nu\nu'}(\mathbf{p}-\mathbf{k}) \equiv  \frac12\left [ V^{11}_\text{ind}(\mathbf{p}-\mathbf{k})+{\rm sgn}(\nu,\nu')\right.\nonumber\\
 \left. \times e^{-i\left[\phi_2(\bp)- \phi_1(\bp)\right ]/2} V^{12}_\text{ind}(\mathbf{p}-\mathbf{k}) e^{i\left[\phi_2(\bk)- \phi_1(\bk)\right ]/2}\right ],
 \end{align}
 where ${\rm sgn}(\nu,\nu) = 1$ and ${\rm sgn}(\nu,-\nu) = -1$.
Finally the number equation is 
$N =\sum_{\nu,\bp} [1 -  \xi_\bp\tanh(E_{\bp,\nu}/2T)/E_{\bp,\nu}  ]/2$
and the BCS ground state energy  is 
\begin{align}
E_{\rm BCS} -\mu N = \frac12\sum_{\nu,\bp}[\xi_\bp  -E_{\bp,\nu} + |\Delta_{\bp,\nu}|^2/2E_{\bp,\nu}].
\end{align}
We note that when the $s$- and $p$-wave order parameters co-exist, their relative phase is important. It cannot be gauged away contrary to the case of a 
single order parameter. The relative phase therefore has physical consequences, and we shall see that it determines whether the system has a time-reversal symmetry or not.
\section{Symmetries and topological properties}
The topological properties of the bi-layer system are determined by its symmetries  and 2D
 dimensionality~\cite{Schnyder2008,Kitaev2009,Altland1997}. 
 Consider first the limit where  the two layers are uncoupled, which corresponds to the layer distance being much larger 
than the range of the induced interaction given by the BEC coherence length, i.e.\ $d\gg \xi_\text{B}$. There is then only particle-hole symmetry 
for each layer, and they each form a topological $p_x\pm ip_y$ superfluid in symmetry class D, which supports chiral edge states. 
Consider now the case when the two layers are brought closer to each other so that they  interact. 
The topological properties and the fate of the edge states  then depend on whether the Cooper pairs 
in the two layers have opposite or the same angular  momentum, corresponding to $(p_x+ip_y)\times(p_x-ip_y)$ or $(p_x+ip_y)\times(p_x+ip_y)$ pairing respectively.  

For $(p_x+ip_y)\times(p_x-ip_y)$ pairing  illustrated in Fig.\ \ref{fig:system} (a), which we 
refer to as the $(+,-)$ case, the system possesses in addition to particle-hole symmetry the time-reversal symmetry: 
\begin{align}
\mathcal T (a_{\mathbf{p} 1},  a_{\mathbf{p}2})\mathcal T^{-1}=(a_{-\mathbf{p}2},  -a_{-\mathbf{p}1}),
\end{align}
which swaps particles in the two layers. Note that this anti-unitary symmetry is different from the usual time-reversal symmetry, which flips the 
spin of the particles. Here, the layer index plays the role of a pseudo-spin. Since $\mathcal T^2=-1$, the  
bi-layer system is then in symmetry class DIII, and its ground state is a $\mathbb{Z}_2$ topological
 superfluid, which supports helical edge modes in analogy with the quantum spin Hall  state~\cite{Kane2005a,Kane2005b,Qi2009}. 
 The counter propagating edge modes in the two layers 
 are related by TRS and protected by Kramers theorem. However, when the layers are sufficiently close together, 
the $s$-wave inter-layer pairing ($\Delta_{12}(\mathbf p)\neq0$) will dominate, and  the system  forms a topologically trivial $s$-wave superfluid. Thus, 
 the edge states must become gapped at some critical inter-layer distance. Without solving the gap equation, one can envision two ways this can happen: either the inter-layer pairing 
 explicitly breaks TRS thereby gapping the edge modes as soon as $\Delta_{12}(\mathbf p)\neq0$, 
 or the inter-layer pairing respects TRS 
 and the edge states become gapped only when the bulk energy gap is closed. By analysing the properties of the inter-layer gap under  
 time-reversal, we find that 
  these two scenarios correspond to $\Delta_{12}(\mathbf{p})$ being imaginary and real respectively.  Our numerical results (see later) show that $\Delta_{12}(\mathbf{p})$ is in fact imaginary and the first scenario describes the physical transition. 

For $(p_x+ip_y)\times(p_x+ip_y)$ pairing illustrated in Fig.\ \ref{fig:system} (b), which we 
refer to as the $(+,+)$ case, the system only has the particle-hole symmetry and  is a $\mathbb{Z}$ topological
 superfluid in class D, which supports chiral edge modes propagating in the same direction in the two layers. When the layer distance is decreased, the possible onset of inter-layer pairing co-existing 
 with the intra-layer pairing will not gap these edge modes as long as the bulk gap remains non-zero, since this pairing does not break any symmetry. However, we shall see later that such a co-existing scenario does not occur for the $(+,+)$ case. Similar to the $(+,-)$ case,  the system ends up in the topologically trivial  inter-layer $s$-wave superfluid for small inter-layer distances. We shall demonstrate below that this happens via  
 a first order phase transition.

The topological $\mathbb{Z}$ and $\mathbb{Z}_2$ invariants of class D and DIII respectively can  be calculated from the two energy bands $E_{\mathbf p,+}$ and $E_{\mathbf p,-}$ 
of the bilayer system~\cite{deLisle2014}. If the two layers are uncoupled,  these bands are degenerate and the invariant  for class D is simply given by the 
sum $C = C_1+C_2$ of the  Chern numbers $C_\sigma$ of each layer, whereas it is given by the difference  $\nu=C_1-C_2\, (\text{mod}\, 2)$ 
for class DIII. This is consistent with the fact that 
the $(+,-)$ state has $C_1=-1,C_2=1$ and is therefore topological in class DIII, whereas it is trivial in class D. Therefore, if the TR symmetry is broken for the $(+,-)$ state by an imaginary $\Delta_{12}(\mathbf p)$ that mixes the two bands, the system is in class D and it is 
no longer topological. 

\section{Edge states}
In this section, we show explicitly how the edge states of   the $(+,-)$ system become gapped with the onset of interlayer $s$-wave pairing $\Delta_{12}(\mathbf{p})$ which is imaginary. We consider the following low-energy hamiltonian in real space,
\begin{align*}
H = \int d^2\mathbf{r} \, \Psi^\dagger(\mathbf{r}) \mathcal{H}(\mathbf{r}) \Psi(\mathbf{r})
\end{align*}
where $\Psi(\mathbf{r}) = (\psi_1, \psi^\dagger_1, \psi_2, \psi^\dagger_2)^T$ and
\begin{widetext}
\begin{align*}
\mathcal{H}(\mathbf{r}) = 
\begin{pmatrix}
-\mu(\mathbf{r}) & \Delta_{11} e^{-i\phi_0}(-\partial_x +i\partial_y) & 0 & \Delta_{12} \\
 \Delta_{11} e^{i\phi_0}(\partial_x +i\partial_y) & \mu(\mathbf{r}) & -\Delta_{12}^\ast & 0 \\
 0 & -\Delta_{12} & -\mu(\mathbf{r}) & \Delta_{11} e^{i\phi_0}(-\partial_x -i\partial_y)  \\
 \Delta_{12}^\ast & 0 & \Delta_{11} e^{-i\phi_0}(\partial_x -i\partial_y) & \mu(\mathbf{r})
 \end{pmatrix}
\end{align*}
\end{widetext}
Assuming that we can apply a local density approximation, we take $\mu(\mathbf{r}) = \mu(r)$ to be positive within the radius $R$, and negative outside. Solutions to the eigenvalue equation $\mathcal{H}(\mathbf{r}) \chi(\mathbf{r})=E \chi(\mathbf{r})$ with definite angular momentum can then be found, and we use the following ansatz in the usual polar coordinates,
\begin{align*}
\chi(\mathbf{r}) = \kappa e^{in \theta}
\begin{pmatrix}
e^{-i\phi/2}[A(r)+iB(r)] \\
e^{i\phi/2}[A(r)-iB(r)] \\
e^{i\phi/2}[C(r)+iD(r)] \\
e^{-i\phi/2}[-C(r)+iD(r)]
\end{pmatrix}
\end{align*}
where $\kappa$ is a normalization constant. The real functions $A,B,C,D$ satisfy a set of 4 coupled equations, and for a large system with tightly confined edge modes, we can find solutions with the energy $\pm E = \pm \sqrt{\left( \Delta_{11} n/R \right)^2 + |\Delta_{12}|^2}$. Here, $n$ is a half-integer related to the angular momentum of the edge state. These solutions requires $\Delta_{12}$ to be real. If it is imaginary, it is possible to show that the edge modes do not acquire a gap. When finding the specific solutions for the edge states, care should be taken to choose the solution that is normalizable and confined to the edge. As an example, consider the physical, positive branch of energies, $+E$. A possible solution is given by $A(r) = D(r) = 0$ and
\begin{align*}
B(r) &= \exp \left\lbrace \frac{1}{\Delta_{11}} \int_0^r \mu(r') dr' \right\rbrace \\
C(r) &= \alpha \cdot B(r)
\end{align*}
with
$$
\alpha= \frac{\Delta_{11} n}{|\Delta_{12}| R} - \sqrt{\left( \frac{\Delta_{11}n}{|\Delta_{12}| R}\right)^2 +1}.
$$
We see that the edge states lowest in energy are localized on both layers when the two gap parameters coexist. The states higher in energy approach the solutions for uncoupled layers, which are only localized on a single layer.

\section{Numerical solution of the gap equation}
 We now numerically solve the 
  gap equations~(\ref{gap3}) along with the number equation at $T=0$. The $(+,-)$ case corresponds to 
 $\phi_2(\mathbf{p})-\phi_1(\mathbf{p}) = \pi - 2\varphi_{\mathbf{p}}$, while the $(+,+)$ case 
   corresponds to  $\phi_2(\mathbf{p})-\phi_1(\mathbf{p}) = \pi $. 
 
 \subsection{$(+,-)$ system}
\begin{figure}[htb]
\centering
\includegraphics[width=\columnwidth]{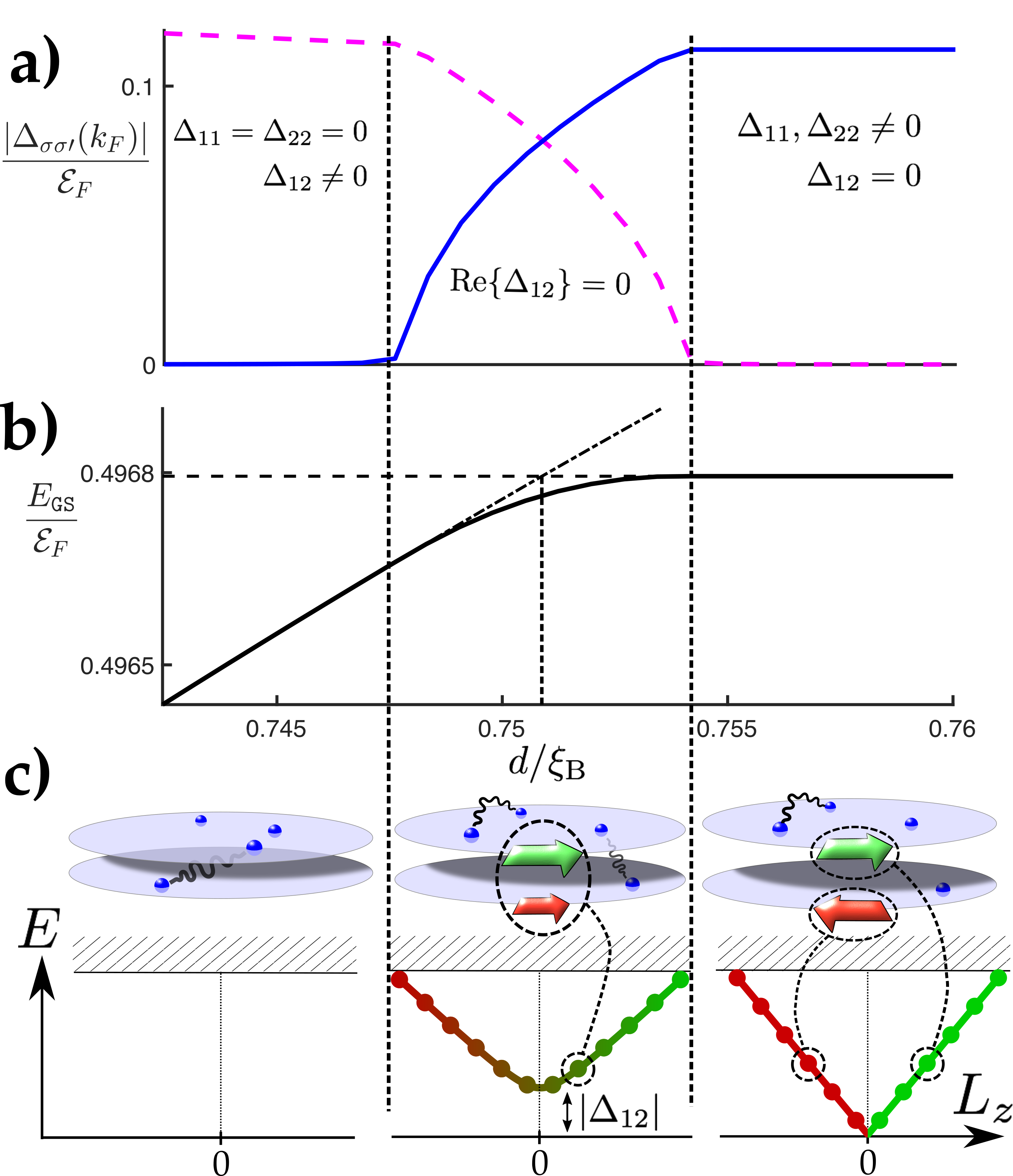}
\caption{(Color online). (a) The magnitude of the  inter-layer $s$-wave pairing (dashed line) and intra-layer $p$-wave pairing (solid line)
as a function of the layer distance for the  $(+,-)$ system. (b) The  corresponding ground state energy per particle   (solid line). The the dashed lines indicate the energy of states with only inter- or intra-layer pairing. The dashed vertical line at $d\simeq0.751\xi_B$ indicates where the two solutions have the same energy. 
(c) The edge modes of the $(+,-)$ system and their spectrum.   For $d\gtrsim 0.754\xi_B$ (right column)   the counter clockwise/clockwise 
 edge modes are localied in  the upper/lower layer. For $ 0.747\xi_B\lesssim d\lesssim 0.754\xi_B$ (middle column),   the low lying edge modes are localised in both layers and they acquire a gap.
 For $ d\lesssim 0.747\xi_B$ (left column), there are no edge modes.}
\label{fig:Z2}
\end{figure}

  In  Fig.~\ref{fig:Z2} (a), we plot the magnitude  of the pairing fields at the Fermi surface  as a function of the layer distance $d$ 
   for the $(+,-)$ system.  Here $k_F=\sqrt{4\pi n_F}$ is the Fermi momentum with $n_F$
the density of fermions in each layer. We have chosen a relatively weak Bose-Fermi coupling strength $g=2\pi a/\sqrt{m_r m_B}$ with 
$k_Fa=0.1$, where  $a$ is the 2D-3D mixed dimensional scattering length~\cite{Nishida2008}. The gas parameter of the BEC is $(n_B a_{B}^3)^{1/3}=0.01$ and the ratio of the Fermi and Bose 
 interparticle distances is $n_F^{1/2} / n_B^{1/3}=0.2$. The energy of the system is plotted in Fig.~\ref{fig:Z2} (b).
For  layer distances $d\gtrsim0.754\xi_B$, there is no inter-layer pairing and 
the two layers are uncoupled each realising a  $p_x\pm ip_y$ topological superfluid. 
The corresponding edge states, illustrated in Fig.~\ref{fig:Z2} (c), propagate in opposite directions in the two layers and are related by the 
TRS operator $\mathcal T$. We have  chosen a circular  boundary with radius $R$ to illustrate the typical geometry formed by the harmonic trap 
in an atomic gas experiment. 
As the layer distance decreases, inter-layer pairing sets in for $d\lesssim0.754\xi_B$ via a second order transition and it co-exists with the intra-layer pairing. 
We find numerically that the inter-layer pairing $\Delta_{12}(\mathbf k)$ is purely imaginary and it therefore breaks TRS. 
The edge modes in the two layers  mix and 
become gapped as illustrated in Fig.~\ref{fig:Z2} (c). More precisely, the 
 dispersion of the edge modes is $E = \sqrt{( \Delta_{11} n/R )^2 + |\Delta_{12}|^2}$, where $|\Delta_{12}|\simeq|\Delta_{12}(0)|$ and 
 $\Delta_{11}(\mathbf p)\simeq\Delta_{11}(p_x+ip_y)$ give the magnitude of the inter- and intra-layer pairing at low momenta, and 
 $n$ is a half-integer  proportional to the angular momentum of the edge state, as seen above.
  The low-energy edge states with small $n$ are hybridised between the two layers; for larger $n$, the edge states become increasingly localized in a single layer, approaching those for the uncoupled layers.
   Finally, for  layer distances $d\lesssim0.747\xi_B$ the intra-layer pairing is completely suppressed by the inter-layer pairing 
   and the system is a topologically trivial 
$s$-wave superfluid with no edge modes. We have not been able to find a numerical solution with a real inter-layer pairing 
co-exisiting with intra-layer pairing, which would preserve TRS and support the gapless edge modes.
While the co-existence region shown here is quite narrow, the width can be tuned by altering the parameters (see below). 
 
  \subsection{$(+,+)$ system}
 For the $(+,+)$ system, our numerical results show that the transition between the topological and trivial phase is first order.
 The transition occurs at the critical layer  distance $d\simeq0.751\xi_B$   when the phases
  with only one type of pairing have the same energy, as indicated by the vertical line in Fig.~\ref{fig:Z2}(b). We do not find numerical solutions with both types of pairing coexisting. Instead, the intra-layer pairing and the associated gapless edge modes disappear and the inter-layer pairing appears abruptly.

 \section{Varying the coherence length}
\begin{figure}[htb]
\centering
\includegraphics[width=\columnwidth]{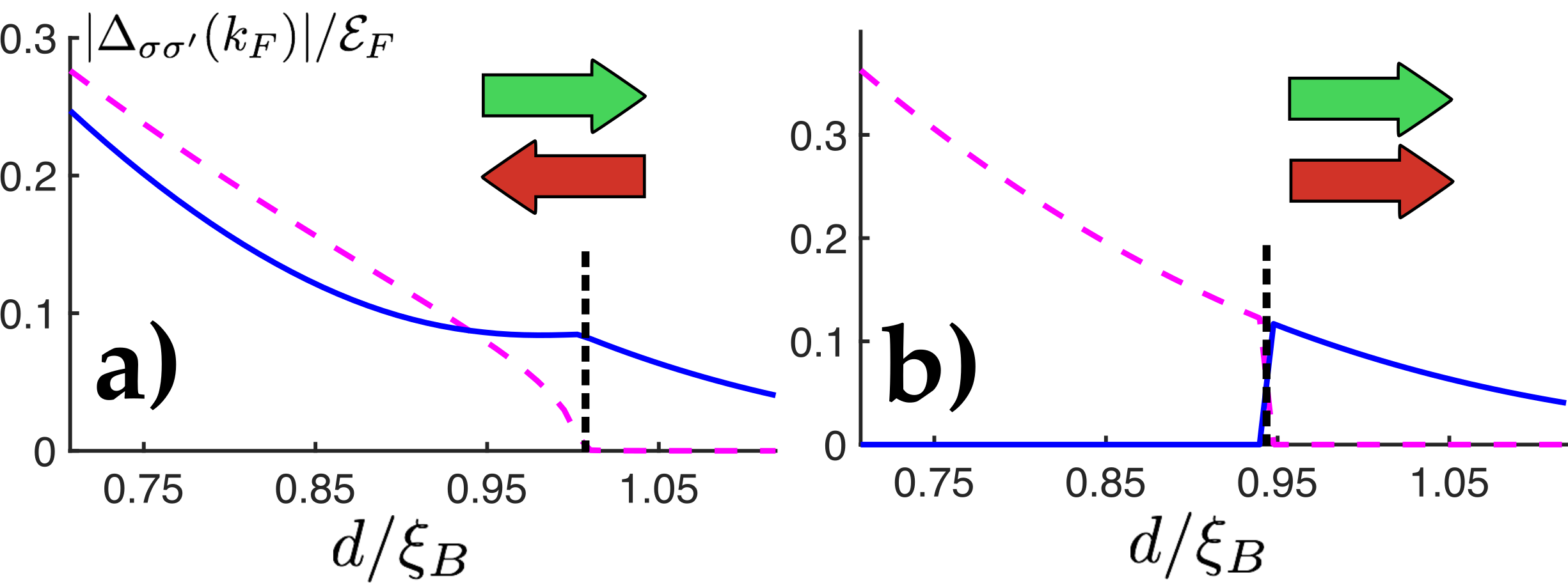}
\caption{(Color online). The intra- (solid line) and inter-layer (dashed line) pairing as a function of the BEC coherence length for the 
$(+,-)$ (a) and $(+,+)$ (b) system. 
}
\label{fig:Varyxi}
\end{figure}
Experimentally, it might be easier to change the BEC coherence length, which determines the range of the induced interaction, by varying $a_B$ using a Feshbach resonance, instead of changing the layer distance. To examine 
this case, we plot in Fig.\ \ref{fig:Varyxi} the magnitudes of the intra- and inter-layer pairings as a function of $\xi_B$ with $k_F a = 0.12$, $k_F d = 1.0$, and $n_F^{1/2} / n_B^{1/3}=0.2$.
The coherence length is varied by changing $a_B$ keeping $n_B$ fixed. 
 For a small $\xi_B$, the two layers are uncoupled forming  the $(+,-)$ or the $(+,+)$ topological superfluid. 
The  $(+,-)$ system undergoes a second order phase transition to a state where intra- and inter-layer pairing co-exist for $\xi_B\gtrsim 1d$. Note that contrary to decreasing the distance 
$d$, the system does not end up in a pure $s$-wave state for large $\xi_B$. The reason is that for a large interaction range, the suppression of 
the $p$-wave channel compared to the $s$-wave channel is small, and intra- and interlayer pairing therefore  co-exist. The $(+,+)$ system on the other hand again undergoes 
a first order transition between the topological and the trivial phases at $\xi_B\sim1.05d$.

\section{Discussion}
All the ingredients in the proposed setup have  been realised experimentally. 
Bose-Fermi mixtures as well as species selective optical potentials  to produce mixed dimensional systems have been 
reported~\cite{Lamporesi2010,McKay2013,Jotzu2015}. It was moreover shown in Ref.~\cite{Wu2016} that the Berezinskii-Kosterlitz-Thouless critical temperature 
for the $p_x\pm ip_y$ superfluid in the present Bose-Fermi setup can be as high as $T_{\rm BKT}= \mathcal{E}_F/16$, which  is  within experimental reach~\cite{Onofrio2017}.
We expect the critical temperature of the phase with $s$-wave pairing to be even higher.
The edge modes can be observed for instance by direct imaging or by the response to an external drive in analogy with topological insulators~\cite{Goldman2013,Tran2017, Goldman2016}.

An intriguing question concerns the robustness of the edge modes beyond mean-field BCS theory. To investigate this, one could analyse the coupling between the edge modes 
forming a Luttinger liquid~\cite{Cenke2006,Tanaka2009}, which is an interesting future project.

\section{Conclusion}
We demonstrated that a  mixed dimensional system consisting of two layers of fermions in  a  BEC is a powerful setup to realise topological superfluids with 
TRS. The induced interaction between the fermions mediated by the BEC  leads to a 
competition between $p$-wave pairing within each layer and $s$-wave pairing between the layers. For large layer separation or short BEC coherence length, intra-layer pairing 
dominates and the system forms a topological superfluid either with or without TRS. In the case of TRS, 
the system goes from a $\mathbb{Z}_2$ topological superfluid to a topologically trivial superfluid via a second order transition where  $s$-wave 
pairing gradually gaps the edge modes. When there is no TRS, the transition from a  $\mathbb{Z}$ topological superfluid to a topologically trivial superfluid is first order. 
These results show how cold atomic gases offer a realistic path to realising topological superfluids with TRS. 

\begin{acknowledgments}
 We wish to acknowledge the support of the Villum Foundation via grant
  VKR023163 and the  Danish Council of Independent Research | Natural Sciences via Grant No. DFF - 4002-00336. We  acknowledge useful discussions with 
 Alexandre Dauphin  and Alessio Recati. 
\end{acknowledgments}

 \bibliographystyle{apsrev4-1}
\bibliography{Ref_bilayer}

\begin{thebibliography}{48}
\bibitem [{foo()}]{footnote}%
  \BibitemOpen
  \href@noop {} {\bibinfo  {journal} {We shall from now on refer to superfluids
  only. It is implicitly understood that this also refers to superconductors.}\
  }\BibitemShut {NoStop}%
\end{thebibliography}

\end{document}